\documentclass[twocolumn]{jpsj3}
\newcommand{\nn}{\nonumber}

\def\runauthor{Online-Journal Subcommittee}

\title{%
Textures and Vortices in $d$-Wave Fermi Condensates in Atomic Gases
}

\author{%
Hiroki~M. \textsc{Adachi}\thanks{E-mail address: hiroki@mp.okayama-u.ac.jp}, 
Yasumasa \textsc{Tsutsumi}, and Kazushige \textsc{Machida}
}

\inst{%
Department of Physics, Okayama University,
Okayama 700-8530
}

\recdate{\today}

\abst{%
Fundamental properties of superfluids with
$d$-wave pairing symmetry are investigated theoretically.
We consider neutral  atomic Fermi gases in a harmonic trap,
the Cooper  pairing being produced by a Feshbach resonance via a $d$-wave interaction channel.
A Ginzburg-Landau (GL) functional is constructed which is symmetry constrained
for five component order parameters (OPs).
We find stable OP textures and vortices for all the three phases which are known to be the
energy minimum of the GL functional; the ferromagnetic, polar and  cyclic phases both
at rest and under rotation. In particular, we touch upon
the stability conditions for a non-Abelian fractional 1/3-vortex 
in the cyclic phase under rotation. It is proposed how to
create the intriguing 1/3-vortex experimentally in atomic gases via optical means.
}

\kword{$d$-wave pairing, superfluids, neutral atomic Fermi gases, vortex, texture, 1/3-vortex
}

\begin{document}
\maketitle

\section{Introduction}

Superfluids with $s$-wave pairing symmetry have been realized by
using a Feshbach resonance of $^6$Li atom gases in 2005\cite{s-wave}
by sweeping a magnetic field.
It is natural to expect that the research front of cold atom gases develops towards realizing condensates
with a higher relative angular momentum of the Cooper pairs.
In fact, much effort from theoretical and experimental sides is now focused on $p$-wave pairing in
$^6$Li  where a $p$-wave Feshbach resonance is already confirmed  to 
exist\cite{p-wave}. Although $p$-wave molecules have been formed\cite{p-wave}, 
experimental evidence of  $p$-wave superfluidity in 
$^6$Li has not been reported yet.

It might be useful to theoretically investigate superfluidity with higher
angular momentum $l$ of the Cooper pair, such as $d$-wave ($l=2$), or
$f$-wave ($l=3$), in neutral Fermi gases 
to stimulate experimental and theoretical works towards this direction.
Hulet has  performed a coupled channels calculation and found 
a Feshbach resonance for $d$-wave channel in $^6$Li\cite{hulet}, and hence
we have a good reason to explore this possibility from a theoretical point of view.
There have already been appearing theoretical studies towards this direction\cite{demler,wu}.

It is known that the pairing symmetry of high T$_c$ cuprates is described by the $d_{x^2-y^2}$ state.
The study of superfluids with $d$-wave symmetry might be important because several strongly correlated
superconductors, or so-called heavy Fermion superconductors belong to unconventional
pairing states, such as $d$, $d+id$, or $f$, etc, including high T$_c$ whose pairing mechanism is still unclear.
Note that in the heavy Fermion superconductor UPt$_3$ a $f$-wave pairing is realized\cite{machida1,machida2,machida3}.
It is also interesting because of the richer physics associated with the many
internal degrees of freedom of the relative orbital angular momentum $l>1$ of the Cooper pairs
 compared to the $l=0$ ($s$) and $l=1$ ($p$) cases.
In particular,  the spatial structure of the order parameter (OP), or the textures and vortices,
which are hallmarks of superfluidity, are intriguing in the confined ultracold atomic gases.

We note that the OP structures of the $d$-wave superfluid of Fermionic atoms and spinor condensates of
Bosonic spin-2 atoms are mathematically similar because both OPs have 5 components.
Whereas the latter has been extensively investigated already\cite{yip,ueda,walter1,walter2}, 
the former has not been studied so far
in connection with neutral atom gases.  
The boundary conditions due to the harmonic trap are different between the two cases in an essential way,
giving rise to novel textures and vortices in the $d$-wave superfluid.
This difference arises because of the different origin of the OP degeneracy 
due to the internal degrees of freedom: The orbital angular momentum lives in real space in
the $d$-wave superfluid unlike the spin degrees of freedom in the Bosonic case
although both are of the same $SO(3)$ symmetry
in a  homogeneous system.

Mermin\cite{mermin} presents a general framework based on Ginzburg-Landau (GL) 
formalism to describe a $d$-wave superfluid in an infinite bulk system.  He exhaustively classifies
the ground state phase diagram into three regions; ferromagnetic (FM), polar (PO) and cyclic (CY) phases, 
depending on the coupling constants
or fourth order coefficients $\beta_1$, $\beta_2$, and $\beta_3$ in the GL functional.
The resulting phase diagram in the ($\beta_1$, $\beta_3$) 
plane normalized by $\beta_2$ is shown in Fig. 1 where the three phases are enumerated:
the ferromagnetic (FM), polar (PO), and cyclic (CY) phases.
The cyclic phase is the most intriguing one among them because 
Fermion superfluids with $s$-wave ($l=0$) or spinless $p$-wave ($l=1$) pairings produced
by a Feshbach resonance via a magnetic sweep, or the spin-1 spinor BEC\cite{ohmi,ho} do not support this phase
which only appears above $l\geq2$.

\begin{figure}
\begin{center}
\includegraphics[width=8cm]{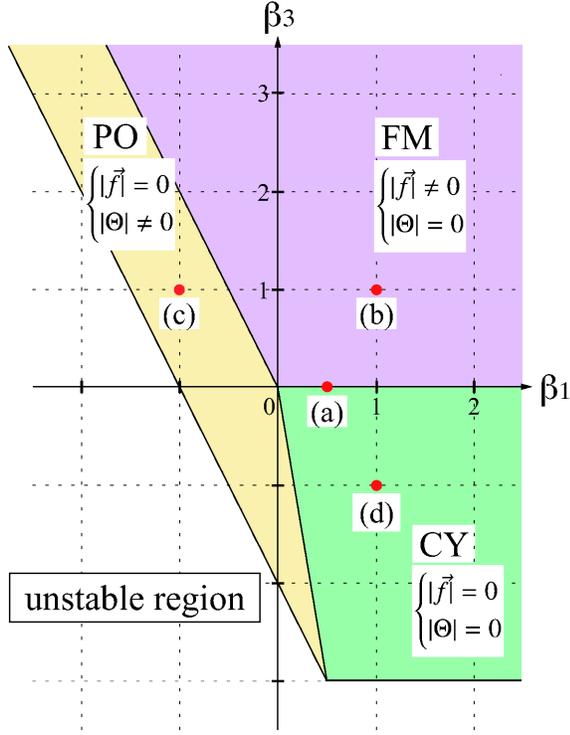}
\end{center}
\caption{(Color online) 
Phase diagram of the stable phases in uniform system
where $\beta_1$ vs $\beta_3$ plane normalized by $\beta_2$
is shown; ferromagnetic (FM), polar (PO) and cyclic (CY) phases.
(a) corresponds to the weak coupling case. (b), (c) and (d) 
correspond to the parameter values adopted in the strong coupling
cases.
}
\end{figure}

The motivation of this paper is to provide the fundamental physical properties of 
a $d$-wave superfluid confined by  a harmonic potential in two-dimensions (2D)
in order to help to identify the $d$-wave nature experimentally,
focusing on the OP spatial structures or textures at rest and vortices under rotation.
We study the most stable textures and vortices for each phase under trapped geometry.
In particular, the  non-Abelian 1/3-vortex 
which is the energetically favored state under external rotation in the cyclic phase
is examined in some details.
The existence of a 1/3-vortex provides a spectroscopic means to characterize 
$d$-wave superfluidity. Mutually non-commutative 1/3-vortices themselves are
already pointed out to be allowed topologically in an $F=2$ spinor BEC by several authors\cite{1/3-1,1/3-2,1/3-3}.
However, there is no serious calculation to examine its stability from energetic point of view 
in $d$-wave Fermion superfluids, which is one of our main purposes in this paper.
As for the n $F=2$ spinor BEC we studied the energetics of 1/3-vortices in our recent paper\cite{jukka}.
The non-Abelian 1/3-vortices might turn out to be useful for quantum computation 
because a spatial arrangement of mutually non-commutative 1/3-vortices can be used for storing information,
namely they can be used as a topologically protected qubit.
This is somewhat similar to the idea that utilizes the Majorana particles bound to the vortex core in chiral
$p+ip$ superconductors\cite{sarma}.
Note that the 1/3-vortex discussed here does not support Majorana zero energy particles.

The arrangement of this paper is following. 
In \S 2, we give a detailed derivation of the GL functional form
relevant to the present trapped systems. The  GL gradient term, which is important 
in the trapped Fermi gases, is derived here. We examine the case in which the GL coefficients
are given by weak coupling estimate. 
This case is situated at the boundary between FM and
CY in the fourth order parameter space mentioned above (see (a) in Fig. 1),
finding that the FM phase is stabilized in
trapped systems over the CY phase. 
In \S 3 we examine the spatial structures, or textures of the three phases when trapped,
and also vortices under rotation for each phase within the strong coupling case. 
In particular, the 1/3-vortex in CY is
discussed in detail because it is unique in CY phase and they are topologically 
interesting due to their non-commutative properties.
In the final section we will devote to conclusions and summary and 
touch on  an experimental method to create a 1/3-vortex
in neutral Fermi atomic gases.
A part of this paper is published in ref. \ref{adachi}.

\section{Formulation and Preliminary}

\subsection {Ginzburg-Landau  functional}
Our arguments are based on GL theory which is general enough to describe a $d$-wave superfluid in terms of
the free energy expanded up to fourth order in the OP. GL theory is valid near the transition temperature $T_c$, 
however, its range of applicability is known to be wider empirically than that. The OP $\Delta({\bf r},\mathbf{\hat{k}})$ for a $d$-wave superfluid is 
spanned by the spherical harmonics $Y_{lm}(\hat{\bf k})$ ($\hat{\bf k}$ is a unit vector in the momentum space.)

\begin{eqnarray}
\Delta({\bf r},\mathbf{\hat{k}})=A_m({\bf r}) Y_{lm}(\hat{\bf k})=\hat{\bf k}_iB_{ij}\hat{\bf k}_j
\end{eqnarray}

\noindent
with $l=2$, $m=-2,\cdots, 2$ and $i,j=1,2,3$. The repeated indices are summed over.
The coefficient $A_m({\bf r})$ is a complex valued functions of ${\bf r}=(x,y)$.
To emphasize the essential points, we consider a two-dimensional system assuming 
that the obtained objects extend uniformly to the third dimension.
Alternatively to $A_m$ we may sometimes use the symmetric traceless 3$\times$3 matrix $B_{ij}$.
The $B$-matrix that has five independent elements  is convenient in
 constructing the GL functional for the OP with $SO(3)$
symmetry in addition to $U(1)$ gauge symmetry. In the following we use these two notations  interchangeably.
The bulk GL free energy functional $f_{bulk}$ is derived by Mermin\cite{mermin} as

\begin{align}
f_{bulk}&=\alpha Tr B^{\ast}B+\beta_1|TrB^2|^2\nn\\
&+\beta_2(TrB^{\ast}B)^2+\beta_3Tr(B^2B^{\ast 2})
\end{align}

\noindent
where $\alpha(T)=\alpha_0(T-T_c)$ and $T_c$ is the transition temperature.
There are three independent fourth order terms $\beta_1$, $\beta_2$ and $\beta_3$.
The trace operation for a matrix is denoted by $Tr$.
After some algebra this bulk energy $f_{bulk}$ is recast into

\begin{align}
f_{bulk}&=\alpha_0(T-T_c)\Sigma_i |A_i|^2+\frac{15}{2}\left\{\beta_2+\frac{1}{3}\beta_3\right. \nn\\
&\left. +(\beta_1+\frac{1}{6}\beta_3)|\Theta|^2-\frac{1}{12}\beta_3|\vec f|^2 \right\} \Sigma_{ij}|A_i|^2|A_j|^2
\end{align}

\noindent
where the orbital singlet pairing amplitude 

\begin{eqnarray}
\Theta=(-1)^iA_iA_{-i}/\Sigma_l|A_l|^2
\end{eqnarray}

\noindent
and  the orbital momentum 

\begin{eqnarray}
{\vec f}=A_i^{\ast}{\vec F}_{ij}A_j/\Sigma_l|A_l|^2
\end{eqnarray}

\noindent
with ${\vec F}$ being the 5$\times$5 spin matrix\cite{isoshima}.

The three phases are characterized by FM ($|\Theta|=0,|{\vec f}|\neq 0$),
CY ($|\Theta|=0,|{\vec f}|=0$), PO ($|\Theta|\neq 0,|{\vec f}|= 0$).
It is clear from eq.~(3) that CY is stable when $\beta_1+\beta_3/6>0$ and $\beta_3<0$, 
while FM and PO phases occupy the other regions of the $(\beta_3, \beta_1)$ parameter space
normalized by $\beta_2$ (see Fig. 1).
The canonical CY phase is of the form $\Delta(\hat{\bf k})=iY_{22}(\hat{\bf k})+\sqrt2 Y_{20}(\hat{\bf k})+iY_{2-2}(\hat{\bf k})$
or in a vectorial notation: $(A_2,A_1,A_0,A_{-1},A_{-2})^T=(i,0,\sqrt2,0,i)^T$. The other cyclic states are obtained
from this by rotations.
The weak coupling estimate for the Fermi sphere ($\beta_2=2 \beta_1$ and $\beta_3=0$)
predicts that the stable phase is on the boundary between the CY and FM phases (see (a) in Fig. 1).

The gradient terms are constructed by considering the possible contractions of 
$(\partial_iB_{jk})^{\ast}$ and $(\partial_lB_{mn})$; namely
there are three independent terms: (1) $(\partial_iB_{jk})^{\ast}(\partial_iB_{jk})$
(2) $(\partial_iB_{jk})^{\ast}(\partial_kB_{ij})$
and (3) $(\partial_jB_{ij})^{\ast}(\partial_kB_{ik})$.
This can be understood from the fact that the decomposition of the two
angular momenta $(L=1)\times (L=2)\rightarrow L=3,2,1$ where the former (latter) corresponds
to $\partial$ (OP).
Then, the gradient energy\cite{note1,note2} can be summed up as 

\begin{align}
f_{grad}&=K_1(\partial_iB_{jk})^{\ast}(\partial_iB_{jk})+K_2(\partial_iB_{jk})^{\ast}(\partial_kB_{ij})\nn\\
&+K_3(\partial_jB_{ij})^{\ast}(\partial_kB_{ik})
\end{align}

\noindent
where $K_1=K$, $K_2=2K$, $K_3=2K$ with $K=2K_0/35$ in the weak coupling approximation
which  also gives $\alpha=2\alpha_0 /15$, $\beta_2=2\beta_1=8\beta_0 /315$, 
OP amplitude $\Delta_0=\sqrt{\alpha_0 /2\beta_0}$
and the coherence length $\xi_0=\sqrt{K_0/ \alpha_0}$. For $\alpha_0\equiv N_F/T_c$, $\beta_0\equiv 7\zeta(3)N_F/16\pi^2k_B^2 T_c^2$, and $K_0\equiv 7\zeta (3)N_F(\hbar v_F)^2 / 48\pi^2 k_B^2 T_c^2$.
We can derive the result of eq. (6) also starting from the $A_m$ representation after simple but tedious calculations,
which is described in Appendix in detail.

In addition to these bulk and gradient terms, we take into account a harmonic trap potential:

\begin{eqnarray}
f_{harmonic}= \omega^2{\bf r}^2 \Sigma_i |A_i|^2.
\end{eqnarray}

\noindent
External rotation may be included  by replacing ${\vec \partial}\rightarrow{\vec \partial}-i(2m/ \hbar){\bf \Omega}\times{\bf r}$
where the rotation axis ${\bf \Omega}\parallel {\bf z}$. Therefore the resulting total GL functional is given by

\begin{eqnarray}
f_{total}=f_{bulk}+f_{grad}+f_{harmonic}+f_{cent}
\end{eqnarray}

\noindent
where $f_{cent}$ is the centrifugal potential\cite{tsutsumi1,tsutsumi2}.
The associated mass current is derived as 

\begin{align}
j_i=\frac{4m}{\hbar}{\rm Im}&[K_1B^{\ast}_{jk}\nabla_iB_{jk}+K_2B^{\ast}_{jk}\nabla_kB_{ij}\nn\\
&+K_3B^{\ast}_{ij}\nabla_kB_{jk}].
\end{align}

The following numerical calculations are done in two-dimensional plane of Cartesian coordinates $(x,y)$.
The coupled GL equations for five components are solved on discretized lattice using 161$\times$161 meshes via a simple iteration.
The OP amplitudes and length are scaled by $\Delta_0$ and $\xi_0$ respectively. The trap frequency $\omega$ is
normalized by $\frac{7}{3}\frac{h/2m}{2\pi \xi_0^2}$ with $m$ the atomic mass. 
We fix $\omega=1/80$ in these units throughout this paper (except for the cases explicitly
stated otherwise). The temperature is fixed to $T=0.8T_c$.
The field of view in the OP contour maps is $30\xi_0\times 30\xi_0$.
The rotation frequency $\Omega$ is normalized by the trap frequency $\omega$.

\subsection{Weak coupling calculation}

Here before discussing the detailed results on each phase in the strong coupling case, 
we briefly touch on the weak coupling case,
which indicates the stable phase on the boundary between FM and CY for an  infinite system as shown in (a)
of Fig. 1, in order to understand the importance of the boundary condition through the gradient terms. 
We perform numerical calculations for a trapped system by solving the GL equations for 
five components. It turns out that FM is energetically favorable over CY in the weak coupling case.
This is because of the boundary condition,
which strongly constraints the possible phase. In fact, FM is more flexible than CY in the sense
that the $\vec f$-vectors can arrange to stabilize the state.
As will be seen later (see Fig. 3 (a)),
the $\vec f$-vectors tend to align along the circumference on the boundary. 
This arrangement is energetically advantageous 
because the four point nodes situated along the $\vec f$ direction, which is
explained later in \S 3.1, tend to lie outside the system, 
leading to maximal gain in the condensation energy.
This is analogous to that in $p$-wave superfluids\cite{tsutsumi1,tsutsumi2}.
This flexible feature is absent in CY because $|{\vec f}|=0$.
Thus under confinement the degeneracy between FM and CY is lifted, the latter being never stabilized over FM
in the weak-coupling limit.

 \begin{figure*}[tttt]
\begin{center}
\includegraphics[width=16cm]{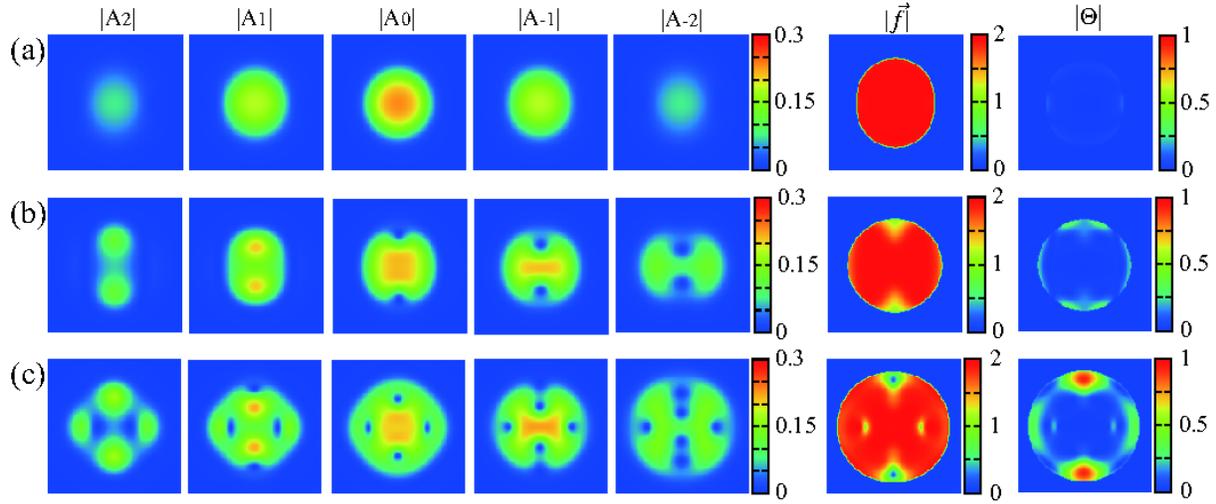}
\end{center}
\caption{(Color online) 
Contour maps of  $|A_2|\sim |A_{-2}|$, $|\vec f|$, and $|\Theta|$ in FM phase.
$\beta_1=1$ and $\beta_3=1$.
(a) $\Omega=0$, (b) $\Omega=0.2$, and (c) $\Omega=0.25$.
}
\end{figure*}

\section{Strong coupling calculation}

\subsection{Ferromagnetic phase}

We start out with the FM phase by choosing the parameter values $\beta_1=1$
and $\beta_3=1$ (see (b) in Fig. 1).
In Fig. 2 we show the contour maps of the  OP magnitude, $|\vec f|$, and $|\Theta|$ for
three cases: (a) $\Omega=0$, (b) $\Omega=0.2$, and (c) $\Omega=0.25$.
It is seen from those figures that at rest the moment vector lies in the
2D plane, pointing to the horizontal $x$-direction mostly and
curving along the circumference regions. There is no $f_z$ component
at rest as seen Fig. 3(a). The non-vanishing components consist exclusively of 
$A_1$, $A_0$, and $A_{-1}$ as seen from Fig. 2(a), which form the in-plane moment
components $f_x$  and $f_y$.
Numerical data show that in our solutions the relation 

\begin{eqnarray}
A_j = A_{-j}^*
\end{eqnarray}

\noindent
holds at every spatial point at rest.
Since 

\begin{eqnarray}
f_z =(2|A_2|^2 +|A_1|^2 -|A_{-1}|^2 -2|A_{-2}|^2)/|\Delta|^2
\end{eqnarray}

\noindent
this leads to $f_z=0$.
As for $\Theta$ it is now given by 

\begin{eqnarray}
\Theta &=(2|A_2|^2 - 2|A_1|^2 +A_0^2)/|\Delta|^2.
\end{eqnarray}

\noindent
Because of the fact that $|A_2|^2 = |A_{-2}|^2 \ll 1$,
$\Theta =(A_0^2 -2|A_1|^2)/|\Delta|^2$.
Thus if $|A_0| = \sqrt{2}|A_1|$, $\Theta=0$.
According to our numerical data the above is nearly satisfied.
Therefore, the obtained phase shown in Fig. 2(a) is genuine ferromagnetic state.

As for the nodal structure of this FM phase,
we start with the most general expression for the $d$-wave OP:

\begin{align}
\Delta(\hat{\bf k})=\sqrt{\frac{5}{8}}&\left[A_2\sqrt3(\hat{k}_x+i\hat{k}_y)^2-A_12\sqrt3(\hat{k}_x+i\hat{k}_y)\hat{k}_z\right. \nn\\ &\left. +A_0\sqrt2(3\hat{k}_z^2-1)+A_{-1}2\sqrt3(\hat{k}_x-i\hat{k}_y)\hat{k}_z\right. \nn\\
&\left. +A_{-2}\sqrt3(\hat{k}_x-i\hat{k}_y)^2 \right].
\end{align}

\noindent
The nodal condition $|\Delta|=0$ is generally satisfied by the four point nodes on
the Fermi sphere.
At the central region where $\vec f$ aligns horizontally the four point nodes
situated at $\theta\sim 90^{\circ}\pm 15^{\circ}$ and $\phi$=$0^{\circ}$
 and $180^{\circ}$ as shown in Fig. 4(a) where the Fermi sphere is described by
 $\theta$ and $\phi$ in the spherical coordinates.
At the circumference region the $\vec f$-vector tends to align so that 
two of the four point nodes are outside of the condensate.
We display an example in Fig 4(b) where the two point  nodes
at $\theta=90^{\circ}$ and $\phi\sim$130$^{\circ}$ and 315$^{\circ}$
 are nearly outside the condensate.
Therefore, it is understood that the curving of the $\vec f$-vectors
in the boundary region is due to the saving of the condensation energy
by excluding the point nodes from the condensate.
This situation is the same as in the $p$-wave condensate in the 
trapped systems\cite{tsutsumi1,tsutsumi2}.


\begin{figure}[tttt]
\begin{center}
\includegraphics[width=8cm]{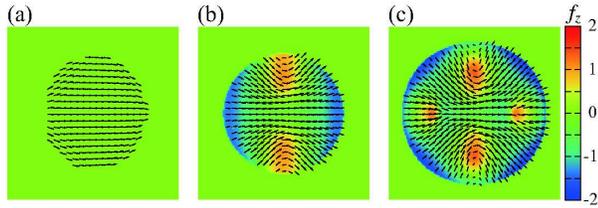}
\end{center}
\caption{(Color online) 
The $f_x$-$f_y$ vectorial patterns are shown, corresponding to (a), (b) and (c) respectively 
in  Fig. 2. The arrow (color code) denotes  $f_x$-$f_y$ ($f_z$) components.
}
\end{figure}

\begin{figure}[tttt]
\begin{center}
\includegraphics[width=8cm]{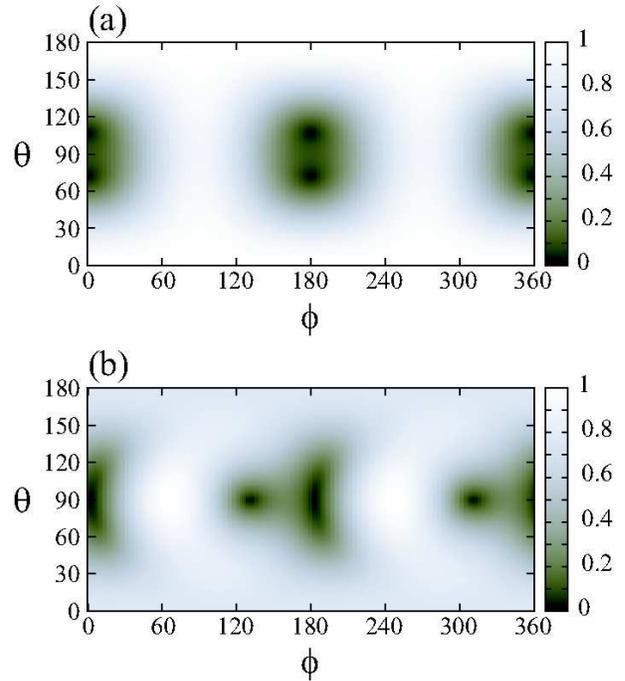}
\end{center}
\caption{(Color online) 
Order parameter amplitude $|\Delta|$ in $k$-space
represented by the spherical coordinates $(\theta, \phi)$
on the unit Fermi sphere.
(a) At the central position and (b) at the upper right corner in 
Fig. 3(a). The darker regions indicate the nodal portion of $|\Delta|$.
}
\end{figure}

Under rotation $\Omega=0.2$ shown in Fig.~2(b) and Fig.~3(b)
the $A_{-1}$ component exhibits phase singularities on the $y$-axis
seen as two dots in Fig.~2(b). Those singularities in $A_{-1}$
are filled by the $A_{1}$ component seen as two yellow points.
Likewise the two singularities in the $A_{-2}$ component are filled by the $A_2$ component.
At those singularities the $\vec{f}$-vectors wind around near those cores
and $f_z$ component is induced as seen from Fig.~3(b).
In addition, the other components such as $A_{\pm 2}$
are induced by rotation (see Fig.~2(b)).
Upon further increasing the rotation to $\Omega=0.25$,
Fig.~2(c) and Fig. ~3(c) show that the two additional similar objects
enter from the $x$ axis, forming a rhombus configuration
and $A_{\pm 2}$ increase further.
Note, however, that the exact four-fold symmetry is not
attained at this rotation frequency.

\subsection{Polar phase}
We now go on discussing the polar (PO) phase
characterized by $|\vec f|=0$ and $|\Theta|\neq 0$.
We take up the parameter values $\beta_1=-1$ and
 $\beta_3=1$ as a representative point for PO as shown in (c) in Fig.~1.
 Figure 5(a) displays the PO phase where $|A_0|\neq 0$ and all the others vanish, correspondingly 
 $|\vec f|=0$  and $|\Theta|=1$  throughout the system.
 This is genuine PO configuration.
 
 Under rotation at $\Omega=0.18$ shown in Fig.~5(b) the two HQVs emerge from the outside along
 the $x$ axis. Namely  while $|A_0|$ remains intact, $A_{\pm 2}$ components mainly
 carry two HQVs. It is seen from Fig.~5(b) that $|A_{- 2}|$ exhibits the suppressions 
 on the $x$ axis, which are compensated by $A_2$ component.
 The OP can be described locally around one of those singular points by 
 the half quantum vortex form: $\Delta({\bf \hat{k}})=\exp(i\varphi/2)
 \exp(-i(\varphi/4)\vec{F}_z)[Y_{2,2}({\bf \hat{k}})+ Y_{2,-2}({\bf \hat{k}})]
 =Y_{2,2}({\bf \hat{k}})+e^{i\varphi}Y_{2,-2}({\bf \hat{k}})$.
 Correspondingly at the core regions the ferromagnetic component is induced 
 (see $|\vec f|$ and $|\Theta|$ in Fig.~5(b)).
 It should be noted here that since there is
 non-vanishing component of $A_0$ at those cores, those vortices are not 
 true half-quantum vortex.
 
 At $\Omega=0.25$ it is seen from Fig.~5(c) those HQV-like objects increase in 
 number which is now four. Moreover a new feature at this
 higher rotation emerges:
 four singular points in $A_0$ are seen in Fig.~5(c). Thus four
 HQV-like objects and four singular cores in $A_0$ coexist in the system,
 each of which are paired up as seen from $|\vec f|$ and $|\Theta|$ in Fig.~5(c).
 Note that those two kinds of objects show only non-vanishing $f_z$
 component. There is no $f_x$ and $f_y$ components throughout the system
because  $A_{\pm 1}$ are completely absent.

\begin{figure*}[hhht]
\begin{center}
\includegraphics[width=16cm]{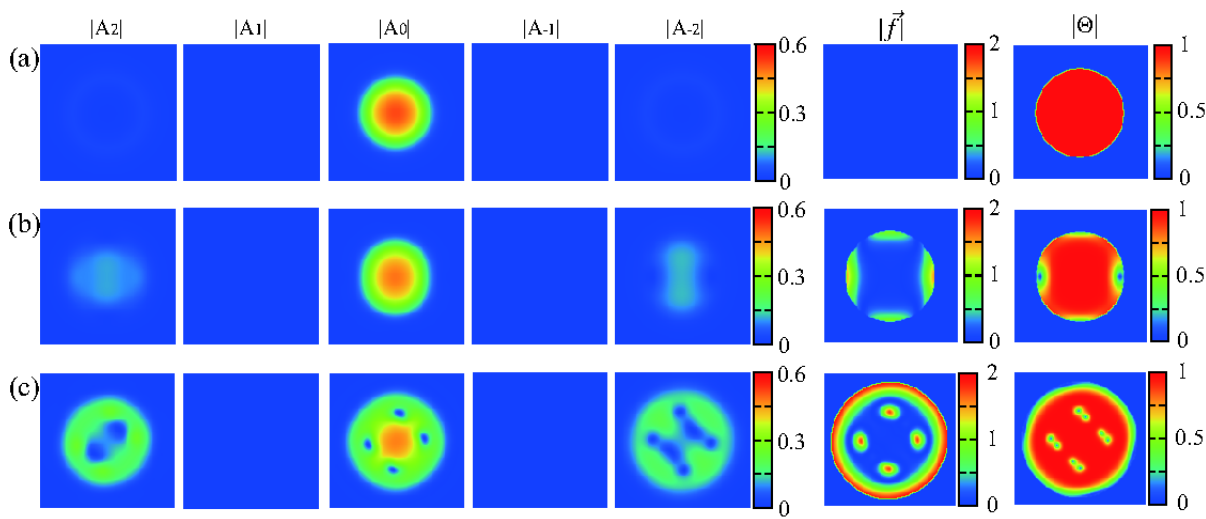}
\end{center}
\caption{(Color online) 
Contour maps of  $|A_2|\sim |A_{-2}|$, $|\vec f|$, and $|\Theta|$ in PO phase.
$\beta_1=-1$ and $\beta_3=1$.
(a) $\Omega=0$, (b) $\Omega=0.18$, and (c) $\Omega=0.25$.
}
\end{figure*}

\subsection{Cyclic phase}
We take up the $\beta_i$ values appropriate for CY;
$\beta_1=1$, $\beta_3=-1$ (see (d) in Fig.~1).
The following results do not depend on these values and are generic in the
cyclic region.
By solving the GL equations, we obtain the stable CY in the presence of a trap. Among various
CY forms derived from the canonical CY described by 
$(i,0,\sqrt2,0,i)^T$, we stabilize the particular CY phase, called CY-z:
$(1,0,0,\sqrt2,0)^T$, which is obtained by 
$\exp(i\cos^{-1}(1/ \sqrt3){\vec F}_y)\exp(-i(\pi/ 4){\vec F}_z)(i,0,\sqrt2,0,i)^T$
through rotations of  the canonical CY.
This CY-z  will be seen to support the 1/3-vortex later.
As seen from  Fig.~6(a) 
CY-z is dominant in the central region because 
$|\vec f|=0$ and simultaneously $|\Theta|=0$. 
The surface region consists of FM or PO phases, which are intermingled.
Note that FM is advantageous in the surrounding region because as mentioned 
the system can gain in the condensation energy by tuning the direction of $\vec f$. 
This OP texture differs completely  from the spinor F=2 BEC where the CY-z extends all the way to
the surface of the cloud\cite{jukka} because of different boundary conditions.
 We emphasize that wherever OP  varies spatially, the gradient coupling terms $f_{grad}$ 
in eq.~(4) inevitably induce other components.

The gap structure of CY consists of 8 point nodes;
In canonical CY: $(i,0,\sqrt2,0,i)^T$ the point nodes are situated at the 8 directions $(\pm1,\pm1,\pm1)$
given by 8 corners of the inscribed cube inside the Fermi sphere. In CY-z the
inscribed cube is rotated so that the $(1,-1,1)$ direction lies now parallel to the $z$ axis.
Thus this CY-z is three-fold symmetric with respect to combined gauge transformation and
rotations about the  $z$ axis.
This is the origin for the possible existence of a 1/3-vortex in CY-z as discussed below.
The mass current given by eq. (9) flows spontaneously circularly around the center
of the trap and gradually decreases farther away from the origin.

\begin{figure*}[tttt]
\begin{center}
\includegraphics[width=16cm]{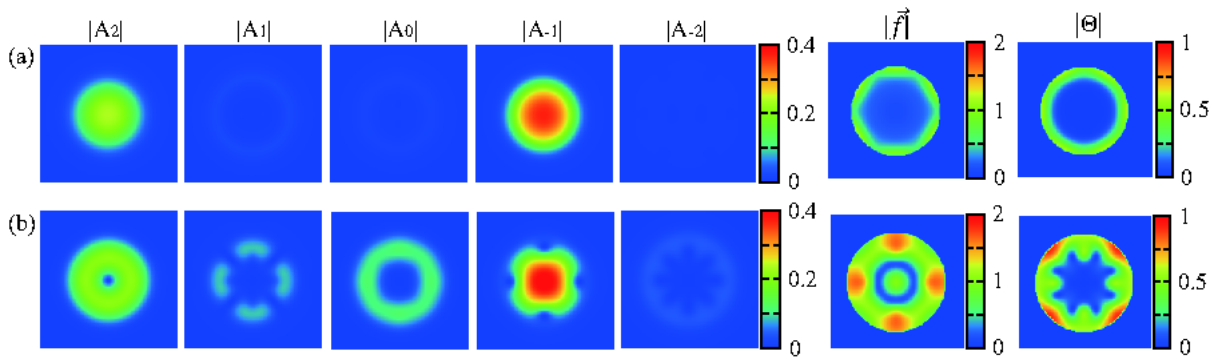}
\end{center}
\caption{(Color online) 
Contour maps of  $|A_2|\sim |A_{-2}|$, $|\vec f|$, and $|\Theta|$ in CY phase.
$\beta_1=1$ and $\beta_3=-1$.
(a) $\Omega=0$ and (b) $\Omega=0.22$.
}
\end{figure*}

By rotating CY-z we can create the 1/3-vortex whose main structure is described by
$(e^{i\varphi},0,0, \sqrt2,0)^T$ in the central region as seen from Fig.~6(b).
Note that at the center the $A_2$ component vanishes whereas the $A_{-1}$  component is non-zero.
Simultaneously the other components are induced around the surface region.
Combination of  the three-fold symmetry around the $z$-axis for CY-z
with an additional gauge transformation leads to the 1/3-vortex form.
Namely, CY-z: $Y_{2,2}(\hat{\bf k})+\sqrt2Y_{2,-1}(\hat{\bf k})$ transforms into
$Y_{2,2}(\hat{\bf k})e^{2i\phi}+\sqrt2Y_{2,-1}(\hat{\bf k})e^{-i\phi}$ under a rotation of an angle $\phi$
around the $z$-axis, which can be rewritten as 
$e^{-i\phi}(Y_{2,2}(\hat{\bf k})e^{3i\phi}+\sqrt2Y_{2,-1}(\hat{\bf k}))=e^{-i\varphi/3}(Y_{2,2}(\hat{\bf k})e^{i\varphi}+\sqrt2Y_{2,-1}(\hat{\bf k}))$
by identifying $\phi=\varphi/3$. After the gauge transformation by $\varphi/3$
this becomes the 1/3-vortex form $(e^{i\varphi},0,0,\sqrt2,0)^T$.
The vortex center is dominated by the $A_{-1}$ component
where the  $A_{2}$ component vanishes due to the phase singularity
at the core as seen from Fig.~6(b). Thus the core of the 1/3-vortex is ferromagnetic.
The nodal structure of the ferromagnetic core region is characterized by a line node on
the equator of the Fermi sphere in addition to point nodes at both poles
because the corresponding basis function is $Y_{2,-1}(\hat{\bf k})\propto \hat{k}_z(\hat{k}_x-i\hat{k}_y)$.

We also calculated the mass current for the 1/3-vortex and the line integral of the
velocity field $\oint v_{\varphi}d\varphi$ along various closed paths around the
center. We find that the line integral yields approximately $(1/ 3)(h/ 2m)$ when circling near the
center, implying 1/3 quantization. However, this conclusion is not exact because the system is inhomogeneous,
 a situation different from the infinite system where the quantization is exact if
the line integral is taken along a path far away from the vortex line.

It is remarkable that this 1/3-vortex is the most stable vortex among various possible ones
which we have solved in order to compare their energies.
The critical rotation frequency is found to be
$\Omega_{cr}=0.21\omega$ at $T=0.8T_c$ beyond which a single 1/3-vortex becomes energetically stable
compared to the CY-z.

\section{Conclusions and summary}
We have examined theoretically textures and vortices for $d$-wave superfluids 
produced by a Feshbach resonance of neutral Fermionic atomic gases.
Our arguments are based on Ginzburg-Landau framework which
is general enough to capture the basic and fundamental
properties of $d$-wave superfluids confined by a harmonic trap
because GL functional is determined by symmetry based arguments; $U(1)$ gauge
symmetry and $SO(3)$ spatial rotation group.

We exhausts all three possible stable phases; ferromagnetic, polar and cyclic phases, 
and show the possible textures and vortices for each phase at rest and under rotation.
We hope that this study helps to identify the $d$-wave superfluidity experimentally.

In particular, the 1/3-vortex characteristic to CY is intriguing 
because this is energetically stable over other possible vortices.
A possible way to realize it experimentally as follows.
The initial preparation of CY-z itself is not difficult where the appropriate ratio $(1:\sqrt{2})$ in
$|A_2|$ and $|A_{-1}|$ is needed. Since these two components are different orbital angular momentum states,
the population in the components can be adjusted using Raman transitions, similarly as 
in F=2 spinor BEC\cite{bigelow1,bigelow2} or by using  RF transition in the presence of Zeeman splitting of the 
5 components under an applied 
field as was done by Hirano's group\cite{hirano}.
Furthermore, once CY-z is prepared, it is possible to imprint the unit phase winding in the 
$A_2$ component using a circular polarized Raman transition\cite{bigelow1,bigelow2}.
It may be difficult to use the optical spoon method to create the vortex, which was utilized to produce
vortices in scalar BECs\cite{madison}.

\section*{Acknowledgements}

We thank T. Ohmi, M. Ozaki, T. Mizushima, M. Kobayashi, T. Kawakami, J.A.M. Huhtam\"{a}ki, T. Hirano and R. Hulet
for useful discussions.

\onecolumn
\section*{Appendix}

In general, the contractions of the tensor $(\partial_i B_{jk})^* (\partial_{\lambda} B_{\mu \nu})$
with the rank six are given by the following three forms:
$(\partial_i B_{jk})^* (\partial_i B_{jk})$, $(\partial_i B_{jk})^* (\partial_k B_{ij})$, and 
$(\partial_j B_{ij})^* (\partial_k B_{ik})$.
The linear combination of these invariants gives rise to the gradient term of the GL functional:
\begin{align}
f_{grad}=&\quad K_1(\partial_i B_{jk})^* (\partial_i B_{jk}) + K_2(\partial_i B_{jk})^* (\partial_k B_{ij}) + K_3(\partial_j B_{ij})^* (\partial_k B_{ik}) \nn\\
&=\quad (K_1+K_2+K_3)\left[ | \partial_1 B_{11} |^2 + | \partial_2 B_{22} |^2 + |\partial_3 B_{33}|^2 \right] \nn\\
&\quad+(2K_1+K_2+K_3)\left[ |\partial_1B_{12}|^2 + |\partial_2B_{23}|^2 + |\partial_3 B_{31}|^2 + |\partial_1B_{31}|^2 + |\partial_2B_{12}|^2 + |\partial_3 B_{23}|^2  \right] \nn\\
&+K_1\bigl[ \left(|\partial_1 B_{22}|^2 + |\partial_2 B_{33}|^2 + |\partial_3 B_{11}|^2 + |\partial_1 B_{33}|^2 + |\partial_2 B_{11}|^2 + |\partial_3 B_{22}|^2\right)  \nn\\
&\quad \quad +2\left(|\partial_1B_{23}|^2 + |\partial_2B_{31}|^2 + |\partial_3 B_{12}|^2 \right) \bigr] \nn\\
&+K_2 \bigl[ ( \partial_1 B_{23})^* (\partial_2 B_{31}) + ( \partial_2 B_{31})^* (\partial_3 B_{12}) + (\partial_3 B_{12})^* ( \partial_1 B_{23}) 
+ ( \partial_1 B_{33 })^* (\partial_3 B_{31} ) + ( \partial_2 B_{11 })^*(\partial_1 B_{12} ) \nn\\
&\quad \quad + (\partial_3 B_{22 })^*(\partial_2 B_{23}) + ( \partial_3 B_{11 })^* (\partial_1 B_{31} ) + ( \partial_1 B_{22 })^*(\partial_2 B_{12} ) 
+ (\partial_2 B_{33 })^*(\partial_3 B_{23}) +C.C. \bigr] \nn\\
&+K_3 \bigl[ (\partial_2 B_{23})^* ( \partial_1 B_{31})+ (\partial_3 B_{31})^*(\partial_2 B_{12} ) + ( \partial_1 B_{12})^*(\partial_3 B_{23}) 
+ ( \partial_1 B_{11 })^* (\partial_3 B_{31} ) + ( \partial_2 B_{22 })^*(\partial_1 B_{12} ) \nn\\
&\quad \quad + (\partial_3 B_{33 })^*(\partial_2 B_{23}) + ( \partial_3 B_{33 })^* (\partial_1 B_{31} ) + ( \partial_1 B_{11 })^*(\partial_2 B_{12} ) 
+ (\partial_2 B_{22 })^*(\partial_3 B_{23}) +C.C.\bigr] .
\label{appendix feB}
\end{align}
On the other hand,
the GL functional can be also written in terms of the OP $A_k$ with $k=-2,-1,0,1,2$. 
In particular, the gradient term is rewritten as 
\begin{align}
f_{g} =- N_F C_2 (\hbar v_F)^2 \frac{1}{3} 
\langle Y_{1,i}^*({\bf \hat{k}}) Y_{1,j}({\bf \hat{k}}) Y_{2,k}^*
({\bf \hat{k}}) Y_{2,l}({\bf \hat{k}}) \rangle_{\bf \hat{k}} (\partial_i A_k)^* (\partial_j A_l)
\end{align}
where $N_F$ is the density of states at the Fermi level.
The coefficient $C_2$ is given by
\begin{align}
C_2 =-\pi k_B T \sum_{\omega_n = - \infty}^{\infty} \frac{1}{4|\omega_n|^3} = -\frac{7\zeta (3)}{4(2\pi k_B T)^2}
\end{align}
where $\omega_n$ is the Matsubara frequency.
Therefore
\begin{align}
f_{g}=\frac{7\zeta (3)N_F(\hbar v_F)^2}{48\pi^2 k_B^2 T^2} 
\langle Y_{1,i}^*({\bf \hat{k}}) Y_{1,j}({\bf \hat{k}}) Y_{2,k}^*({\bf \hat{k}}) Y_{2,l}({\bf \hat{k}}) \rangle_{\bf \hat{k}} (\partial_i A_k)^* (\partial_j A_l).
\end{align}
Near $T \rightarrow T_c$, the above GL functional becomes 
\begin{align}
f_{grad} =K_0 \langle Y_{1,i}^*({\bf \hat{k}}) Y_{1,j}({\bf \hat{k}}) Y_{2,k}^*({\bf \hat{k}}) Y_{2,l}({\bf \hat{k}}) \rangle_{\bf \hat{k}} (\partial_i A_k)^* (\partial_j A_l)
\end{align}
where we define $K_0 \equiv 7\zeta (3)N_F(\hbar v_F)^2 / 48\pi^2 k_B^2 T_c^2$.
Among various combinations of the above indices $i,j,k,l$, 
the non-zero contributions are classified into three cases (A), (B), and (C):\\
(A) $i=j, k=l$
\begin{align}
f_{grad}  =& K_0 \langle |Y_{1,i}|^2 |Y_{2,k}|^2 \rangle |\partial_i A_k|^2 \nn\\
=& K_0 \bigl[ \langle |Y_{1,0}|^2 |Y_{2,0}|^2 \rangle |\partial_0 A_0|^2 
+ \langle |Y_{1,0}|^2 |Y_{2,1}|^2 \rangle \left( |\partial_0 A_1|^2 + |\partial_0 A_{-1}|^2 \right) \nn\\
&+ \langle |Y_{1,0}|^2 |Y_{2,2}|^2 \rangle \left( |\partial_0 A_2|^2 + |\partial_0 A_{-2}|^2 \right) 
+ \langle |Y_{1,1}|^2 |Y_{2,0}|^2 \rangle \left( |\partial_+ A_0|^2 + |\partial_- A_0|^2 \right) \nn\\
&+ \langle |Y_{1,1}|^2 |Y_{2,1}|^2 \rangle \left( |\partial_+ A_1|^2 + |\partial_+ A_{-1}|^2 + |\partial_- A_1|^2 + |\partial_- A_{-1}|^2 \right) \nn\\
&+ \langle |Y_{1,1}|^2 |Y_{2,2}|^2 \rangle \left( |\partial_+ A_2|^2 + |\partial_+ A_{-2}|^2 + |\partial_- A_2|^2 + |\partial_- A_{-2}|^2 \right)  \bigr]
\end{align}
where $\langle |Y_{1,0}|^2 |Y_{2,0}|^2 \rangle =11/7$, etc.
The other cases (B) $i=j\pm 1, k=l\mp 1$ and (C) $i=j\pm 2, k=l\mp 2$ are treated similarly.
By summing up those terms (A), (B), and (C) we obtain
\begin{align}
f_{grad}=&\frac{K_0}{7}\Bigl[ 11|\partial_0 A_0|^2 + 9 \left( |\partial_0 A_1|^2 + |\partial_0 A_{-1}|^2 \right) + 3 \left( |\partial_0 A_2|^2 + |\partial_0 A_{-2}|^2 \right) 
+ 5 \left( |\partial_+ A_0|^2 + |\partial_- A_0|^2 \right) \nn\\
&\quad + 6 \left( |\partial_+ A_1|^2 + |\partial_+ A_{-1}|^2 + |\partial_- A_1|^2 + |\partial_- A_{-1}|^2 \right) 
+ 9 \left( |\partial_+ A_2|^2 + |\partial_+ A_{-2}|^2 + |\partial_- A_2|^2 + |\partial_- A_{-2}|^2 \right)  \nn\\
&\quad + \bigl\{ \sqrt{3} [ (\partial_0 A_1)^* (\partial_+ A_0) + (\partial_- A_0)^* (\partial_0 A_{-1}) 
- (\partial_0 A_0)^* (\partial_+ A_{-1}) - (\partial_- A_1)^* (\partial_0 A_0) ] \nn\\
&\quad\quad + 3\sqrt{2} [ (\partial_0 A_2)^* (\partial_+ A_1) + (\partial_- A_{-1})^* (\partial_0 A_{-2}) 
- (\partial_0 A_{-1})^* (\partial_+ A_{-2}) - (\partial_- A_2)^* (\partial_0 A_1) ] \nn\\
&\quad\quad + 6 (\partial_- A_1)^* (\partial_+ A_{-1}) + 2\sqrt{6} [ (\partial_- A_2)^* (\partial_+ A_0) + (\partial_- A_0)^* (\partial_+ A_{-2}) ]  +C.C. \bigr\}\Bigr] .
\end{align}
This is further rewritten as 
\begin{align}
f_{grad}=&\frac{K_0}{7}\Biggl[11|\partial_z A_0|^2 + 9 \left( |\partial_z A_1|^2 + |\partial_z A_{-1}|^2 \right) + 3 \left( |\partial_z A_2|^2 + |\partial_z A_{-2}|^2 \right) 
+ 5 \left( |\partial_xA_0|^2 + |\partial_yA_0|^2 \right) \nn\\
&\quad + 6 \left( |\partial_xA_1|^2 + |\partial_yA_1|^2 + |\partial_xA_{-1}|^2 + |\partial_yA_{-1}|^2 \right) 
+ 9 \left( |\partial_xA_2|^2 + |\partial_yA_2|^2 + |\partial_xA_{-2}|^2 + |\partial_yA_{-2}|^2 \right)  \nn\\
&\quad+ \Biggl\{ \frac{\sqrt{6}}{2}[ (\partial_z A_0)^* \{(\partial_x-i\partial_y) A_{-1}\} + \{(\partial_x+i\partial_y) A_0\}^* (\partial_z A_{-1}) \nn\\
&\quad\quad\quad\quad -(\partial_z A_1)^* \{(\partial_x-i\partial_y) A_0\} - \{(\partial_x+i\partial_y) A_1\}^* (\partial_z A_0) ] \nn\\
&\quad\quad+ 3[ (\partial_z A_{-1})^* \{(\partial_x-i\partial_y) A_{-2}\} + \{(\partial_x+i\partial_y) A_{-1}\}^* (\partial_z A_{-2}) \nn\\
&\quad\quad\quad -(\partial_z A_2)^* \{(\partial_x-i\partial_y) A_1\} - \{(\partial_x+i\partial_y) A_2\}^* (\partial_z A_1) 
- \{(\partial_x+i\partial_y) A_1\}^* \{(\partial_x-i\partial_y) A_{-1}\}] \nn\\
&\quad\quad- \sqrt{6}[ \{(\partial_x+i\partial_y) A_2\}^* \{(\partial_x-i\partial_y) A_0\} + \{(\partial_x+i\partial_y) A_0\}^* \{(\partial_x-i\partial_y) A_{-2}\} ]  +C.C. \Biggr\} \Biggr]
\end{align}
where $\partial_0=\partial_z$ and $\partial_{\pm }=\mp (\partial_x \mp i\partial_y)/\sqrt{2}$.
After rearranging it, we finally obtain 
\begin{align}
f_{grad}
=&\frac{K_0}{7}\Bigl\{ 11\left[ \left| \partial_x A_{3x^2-r^2} \right|^2 + \left| \partial_y A_{3y^2-r^2} \right|^2 + |\partial_z A_{3z^2-r^2}|^2 \right] \nn\\
&\quad +9 \left[ |\partial_xA_{xy}|^2 + |\partial_yA_{yz}|^2 + |\partial_z A_{zx}|^2 + |\partial_xA_{zx}|^2 + |\partial_yA_{xy}|^2 + |\partial_z A_{yz}|^2  \right] \nn\\
&\quad +3 [ |\partial_xA_{yz}|^2 + |\partial_yA_{zx}|^2 + |\partial_z A_{xy}|^2 
+ \left| \partial_x A_{y^2-z^2} \right|^2 + \left| \partial_y A_{z^2-x^2} \right|^2 + |\partial_z A_{x^2-y^2}|^2 ] \nn\\
&\quad +3 [ ( \partial_x A_{yz})^* (\partial_y A_{zx}) + ( \partial_y A_{zx})^* (\partial_z A_{xy}) + (\partial_z A_{xy})^* ( \partial_x A_{yz}) \nn\\
&\quad\quad + (\partial_y A_{yz})^* ( \partial_x A_{zx})+ (\partial_z A_{zx})^*(\partial_y A_{xy} ) + ( \partial_x A_{xy})^*(\partial_z A_{yz}) + C.C. ] \nn\\
&\quad -2\sqrt{3} [ ( \partial_x A_{3y^2-r^2})^* (\partial_z A_{zx} ) + ( \partial_y A_{3z^2-r^2} )^*(\partial_x A_{xy} ) + ( \partial_z A_{3x^2-r^2})^*(\partial_y A_{yz}) \nn\\
&\quad\quad\quad + ( \partial_z A_{3y^2-r^2})^* (\partial_x A_{zx}  ) + ( \partial_x A_{3z^2-r^2} )^* ( \partial_y A_{xy}) + ( \partial_y A_{3x^2-r^2})^*(\partial_z A_{yz}  ) + C.C.] \Bigr\}
\end{align}
where we transform the basis of $A$ from the $z$ component of the orbital angular momentum to the direction of the rectangular coordinates.

In order to confirm that eq. (22)  is equivalent to the gradient form eq. \eqref{appendix feB} written in terms of $B$,
we first note that the transformation between $B$ and $A$ is given as 
\begin{align}
\begin{split}
&A_{3z^2-r^2}=\sqrt{\frac{1}{5}} B_{33}, \ A_{3x^2-r^2}=\sqrt{\frac{1}{5}} B_{11}, \ A_{3y^2-r^2}=\sqrt{\frac{1}{5}} B_{22}, \\
& A_{x^2-y^2}=\sqrt{\frac{1}{15}} \left( B_{11} - B_{22} \right), \ A_{y^2-z^2}=\sqrt{\frac{1}{15}} \left( B_{22} - B_{33} \right), \
 A_{z^2-x^2}=\sqrt{\frac{1}{15}} \left( B_{33} - B_{11} \right) \\
&A_{xy}=\sqrt{\frac{4}{15}} B_{12}, \  A_{yz}=\sqrt{\frac{4}{15}} B_{23}, \  A_{zx}=\sqrt{\frac{4}{15}} B_{31} .
\end{split}
\end{align}
By substituting $B$ in place of $A$, we obtain
\begin{align}
f_{grad}
=& \quad \frac{2}{7}K_0 \left[ | \partial_1 B_{11} |^2 + | \partial_2 B_{22} |^2 + |\partial_3 B_{33}|^2 \right] \nn\\
&+\frac{12}{35}K_0 \left[ |\partial_1B_{12}|^2 + |\partial_2B_{23}|^2 + |\partial_3 B_{31}|^2 + |\partial_1B_{31}|^2 + |\partial_2B_{12}|^2 + |\partial_3 B_{23}|^2  \right] \nn\\
&+\frac{2}{35}K_0 \bigl[ \left(|\partial_1 B_{22}|^2 + |\partial_2 B_{33}|^2 + |\partial_3 B_{11}|^2 + |\partial_1 B_{33}|^2 + |\partial_2 B_{11}|^2 + |\partial_3 B_{22}|^2\right)  \nn\\
&\quad \quad +2\left(|\partial_1B_{23}|^2 + |\partial_2B_{31}|^2 + |\partial_3 B_{12}|^2 \right) \bigr] \nn\\
&+\frac{4}{35}K_0 \bigl[ ( \partial_1 B_{23})^* (\partial_2 B_{31}) + ( \partial_2 B_{31})^* (\partial_3 B_{12}) + (\partial_3 B_{12})^* ( \partial_1 B_{23}) \nn\\
&\qquad+ ( \partial_1 B_{33 })^* (\partial_3 B_{31} ) + ( \partial_2 B_{11 })^*(\partial_1 B_{12} ) + (\partial_3 B_{22 })^*(\partial_2 B_{23}) \nn\\
&\qquad+ ( \partial_3 B_{11 })^* (\partial_1 B_{31} ) + ( \partial_1 B_{22 })^*(\partial_2 B_{12} ) 
+ (\partial_2 B_{33 })^*(\partial_3 B_{23}) +C.C. \bigr] \nn\\
&+\frac{4}{35}K_0 \bigl[ (\partial_2 B_{23})^* ( \partial_1 B_{31})+ (\partial_3 B_{31})^*(\partial_2 B_{12} ) + ( \partial_1 B_{12})^*(\partial_3 B_{23}) \nn\\
&\qquad+ ( \partial_1 B_{11 })^* (\partial_3 B_{31} ) + ( \partial_2 B_{22 })^*(\partial_1 B_{12} ) + (\partial_3 B_{33 })^*(\partial_2 B_{23}) \nn\\
&\qquad+ ( \partial_3 B_{33 })^* (\partial_1 B_{31} ) + ( \partial_1 B_{11 })^*(\partial_2 B_{12} ) 
+ (\partial_2 B_{22 })^*(\partial_3 B_{23}) +C.C.\bigr] 
\end{align}
where we use the traceless property of $B$.
This is indeed the gradient GL functional eq. \eqref{appendix feB} written in terms of $B$ above.
The coefficients are 

\begin{eqnarray}
2K_1=K_2=K_3=(4/35)K_0.
\end{eqnarray}


\begin{thebibliography}{99}
\bibitem{note2}
We also note that in the $F=2$ spinor BEC the gradient term is uniquely given by
$(\partial_iB_{jk})^{\ast}(\partial_iB_{jk})$ because the partial derivative operators
act on real space while the OP lives in spin space. See ref. \ref{jukka}.
\end{thebibliography}
\end{document}